\documentclass[sigconf]{acmart}

\usepackage{booktabs} % For formal tables
\usepackage[utf8]{inputenc}

\setcopyright{acmcopyright}
\acmPrice{15.00}
\acmDOI{10.1145/3230599.3230602}
\acmISBN{978-1-4503-6543-7/18/06}

\acmConference[CERI'18]{5th Spanish Conference in Information Retrieval}{June 2018}{Zaragoza, Spain}
\acmBooktitle{CERI'18: 5th Spanish Conference in Information Retrieval, June 2018, Zaragoza, Spain}
\acmYear{2018}
\copyrightyear{2018}

\acmArticle{2}
\acmPrice{15.00}

\begin{document}
\title{About BIRDS project (Bioinformatics and Information Retrieval Data Structures Analysis and Design)}

\author{Guillermo de Bernardo}
\affiliation{%
  \institution{Enxenio S.L. / Universidade da Coru\~na. CITIC}
  \city{A Coru\~na}
  \state{Spain}
  \postcode{43017-6221}
}
\email{gdebernardo@enxenio.es / gdebernardo@udc.es}

\author{Susana Ladra}
\orcid{1234-5678-9012}
\affiliation{%
  \institution{Universidade da Coru\~na. CITIC / Database Laboratory}
  \city{A Coru\~na}
  \state{Spain}
}
\email{sladra@udc.es}

% The default list of authors is too long for headers.
\renewcommand{\shortauthors}{de Bernardo and Ladra}

%[G]Evitar solapar titulos
%\renewcommand{\shorttitle}{About BIRDS (Bioinformatics and Information Retrieval Data Structures Analysis and Design)}
%[G]Alternativa 2 para evitar solapar titulos
\renewcommand{\shorttitle}{About BIRDS project}

\begin{abstract}
BIRDS stands for ``Bioinformatics and Information Retrieval Data Structures analysis and design'' and is a 4-year project (2016--2019) that has received funding from the European Union's Horizon 2020 research and innovation programme under the Marie Sklodowska-Curie grant agreement No 690941. 

The overall goal of BIRDS is to establish a long term international network involving leading researchers in the development of efficient data structures in the fields of Bioinformatics and Information Retrieval, to strengthen the partnership through the exchange of knowledge and expertise, and to develop integrated approaches to improve current approaches in both fields. The research will address challenges in storing, processing, indexing, searching and navigating genome-scale data by designing new algorithms and data structures for sequence analysis, networks representation or compressing and indexing repetitive data.

BIRDS project is carried out by 7 research institutions from Australia (University of Melbourne), Chile (University of Chile and University of Concepción), Finland (University of Helsinki), Japan (Kyushu University), Portugal (Instituto de Engenharia de Sistemas e Computadores, Investigação e Desenvolvimento em Lisboa, INESC-ID), and Spain (University of A Coruña), and a Spanish SME (Enxenio S.L.). It is coordinated by the University of A Coruña (Spain).

\end{abstract}

%
% The code below should be generated by the tool at
% http://dl.acm.org/ccs.cfm
% Please copy and paste the code instead of the example below.
%



\maketitle

\section{Introduction}

Nowadays computers are used to process huge amounts of information. For example, a major search engine processes tens of thousands of searches per second, and more than 4 terabytes of Internet traffic is generated in 1 minute. This volume of information is expected to be multiplied by a factor of 10 in the following 10 years. We are just beginning to be capable of managing this huge amount of information. However, we are not prepared to address the big challenge posed by the explosion of data coming from the DNA sequencing.

Thanks to the falling cost of genome sequencing, biological data doubles every six months. Recent studies estimate that there will be up to 2 billion human genomes sequenced by 2025, increasing five orders of magnitude in 10 years. This increment can be even greater; for instance, for understanding cancer it will be necessary to sequence thousands of cells of the same tumor.

Processing the amount of information that will make personalized medicine a reality, requires a new approach. A new class of data structures has recently been developed to address the new challenges in storing, processing, indexing, searching and navigating biological data. Similar tasks have been also tackled by researchers in the information retrieval community. Synergies of researchers from both fields can lead to new efficient approaches to improve the technology used for analysis of genome-scale data. Only by being capable of processing large amounts of information we will be able to understand diseases or the development, behavior and evolution of species.

The overall goal of BIRDS is to establish a long term international network involving leading researchers in bioinformatics and information retrieval from four different continents, to strengthen the partnership through the exchange of knowledge and expertise, and to develop integrated approaches to improve current ones in both fields.

\subsection{Objectives}
The specific objectives for the project are:
\begin{enumerate}
\item To improve the scientific excellence of the participants in the design of advanced data structures and algorithms for the fields of bioinformatics and information retrieval, taking advantage of the synergies from those two communities, with the aim of improving their position and competitiveness at the international level.

\item To reinforce existing collaborations and to create new collaborations among all the participants, both between EU and international partners and especially among EU partners, with the aim of developing long-term research partnerships.

\item To increase the number of new researchers being attracted to the field and to improve the education of PhD candidates and postdocs.

\item To transfer the knowledge generated to market and create new innovative business opportunities for SMEs in Europe.
\end{enumerate}

\subsection{Work Packages}
To reach these objectives, BIRDS is structured in six work packages.
Three of the work packages focus on distinct but highly-related research tasks, each with its particular set of problems and challenges:
\vspace{-2mm}
\begin{itemize}
\item Research Task 1: Algorithms for Sequence Analysis.
\item Research Task 2: Compression and Indexing Techniques for Repetitive Data.
\item Research Task 3: Data Structures and Algorithms for Network Analysis.
\end{itemize}
In addition to these work packages, three more are included related to the management and coordination of the project, the dissemination and training activities, and one work package devoted to integrating all research results for their transfer to market and for exploiting them where possible.

We describe these work packages more in detail as follows:

\noindent{\bf WP1: Coordination and Management.}
The goals of this work package are to coordinate all administrative procedures, to regularly monitor and assess the progress of the project, and to document and distribute all relevant technical discussions and decisions between partners. All beneficiaries have been totally involved in the work carried out in this work package.

\noindent{\bf WP2: Algorithms for Sequence Analysis.}
The goals of this work package are to share the expertise on biological sequence analysis and the latest techniques for information retrieval tasks related to string analysis, to make new contributions in sequencing analysis to advance in both fields, and to identify areas of future collaborative research between the partners in this research task. 
For instance, to understand the features, functions, structure and evolution of the DNA, RNA or protein sequences, it is necessary to address several computational tasks, such as pairwise alignment, multiple sequence alignment, mutations or genetic diversity finding, phylogenetic tree construction, etc. There exist already compact data structures and algorithms that successfully solve these problems in the Information Retrieval field, so it is expected they can improve bioinformatics tasks. In addition, it will be studied if bioinformatics common pattern matching algorithms can be used in Information Retrieval problems.

\noindent{\bf WP3: Compression and Indexing Techniques for Repetitive Data.}
The goals of this work package are to share the expertise on compressing and indexing repetitive collections of data in different fields, to make new contributions in compression and indexing techniques for repetitive data, and to identify areas of future collaborative research between the partners in this research. 
The idea that is exploited in this work package is that in general, collections of genomes are highly repetitive. This characteristic is due to two different reasons. One of them is caused by the use of high-throughput DNA sequencing technologies, which fragment a genome in multiple reads, where one portion of the genome is contained in several reads. The other one is that there exist highly repetitive data when storing large sets of individual genomes of organisms from the same species. Hence, it seems promising to use those techniques designed for compressing/indexing repetitive or versioned data. However, some compression techniques or indexes for repetitive collections are not competitive for DNA collections, such as compressed suffix trees. Thus, in this work package we study compression techniques and indexes for large genomic collections, especially those based on LZ77 or grammars.

\noindent{\bf WP4: Data Structures and Algorithms for Network Analysis.}
The goals of this work package are to share the expertise of UDEC, UCHILE and UDC on information retrieval networks and the expertise of INESCID and UH on biological networks, to make new contributions to network representation and analysis in order to progress in bioinformatics and information retrieval fields, and to identify areas of future collaborative research between the partners in this research task.
In this work package biological data is seen not just as one-dimensional strings, but as trees, graphs or more complex networks. This is required for analyzing the three-dimensional folding of proteins or the RNA secondary structure, as well as using more advanced algorithms based in graphs, such as de Bruijn graphs. Some advanced data structures and algorithms have been already useful for complex network analysis problems. In the field of Bioinformatics, for instance, to compute path-based kernels in common metabolic reactions databases, searching for common sub-trees in a collection of phylogenetic trees, or modelling and predicting regulatory network. In the case of Information Retrieval, to represent and navigate Web and social graphs, compressing and indexing labelled trees, etc. In this work package we characterize and exploit similarities between biological and information retrieval networks to advance in both fields.
More concretely, we show now the progress achieved at each task of this work package:

\noindent{\bf WP5: Research Integration, Benchmarking and Evaluation.}
The goals of this work package are to evaluate the viability of the research results obtained from WP2, WP3 and WP4 over real scenarios and to transfer to market and commercialize, if possible, the research results obtained in those work packages.
The tasks of this work package include integration of research results, generating prototypes, testing and evaluating those prototypes and demonstrations.

\noindent{\bf WP6: Dissemination and Demonstration Activities.}
The goals of this work package are to disseminate the findings of the project, to create a network of external researchers that also contribute to the project, to attract and train new researchers in the topic of the project, and to engage and increase the understanding of the non-specialist public in the field.

\section{Project Implementation}

\subsection{Status of each research line}
\noindent{\bf Algorithms for Sequence Analysis.}
The most important results on this work package are the advances in the field of Bioinformatics, where a new time-optimal algorithm for the subproblem “contig assembly” of the genome assembly problem was proposed  \cite{CPM17}. In addition, there were also some advances on the variation calling problem, which allows detecting mutations of an individual with respect to a reference genome.

In addition to continuing the research on the lines opened during the secondments of this first reporting period, there are some non-addressed topics that are still relevant and will be studied during the second reporting period. One of them consists in the transcript assembly problem, which has not been totally solved yet. By solving this problem one can better characterize diseases such as cancer, or some RNA virus, such as HIV. This will be the focus of one PhD thesis, which will be co-advised by staff from UDC and UH. 

\noindent{\bf Compression and indexing techniques for repetitive data.}
There were significant advances, combining the knowledge of the partners on both fields of information retrieval and bioinformatics, as collaborations within the project generate important results, such as:
\begin{itemize}
\item New universal indexes for highly repetitive document collections~\cite{IS16}%(Information Systems 2016)
\item Efficient and compact representations of some non-canonical prefix-free codes~\cite{SPIRE16-PF}%(SPIRE 2016).
\item New compressed representations of sequences with rank/select support~\cite{JDA16, DCC18-RC} %Cal é o segundo DCC???(Journal of Discrete Algorithms 2016, and two papers at DCC 2018) 
\item Efficient representation of trajectories~\cite{SPIRE16-TRA,SPIRE17-TRA} and event sequences~\cite{DCC18-EV}.
% -	Efficient representation of trajectories (SPIRE 2016, SPIRE 2017) and event sequences (DCC 2018).
\item New succinct data structures for self-indexing ternary relations~\cite{JDA17-TER} and spatio-temporal data~\cite{PACIS16}
%-	New succinct data structures for self-indexing ternary relations (Journal of Discrete Algorithms 2016) and spatio-temporal data (PACIS 2016)
\end{itemize}
In addition, there have been some important advances on new compression and indexing techniques for repetitive data using LZ77-like parsings, which have not been published yet, but they are expected to be published during the second reporting period. As they achieve high compression, they will be an important contribution to the field, with applications on both bioinformatics and information retrieval.

During the first reporting period there were very significant advances in this research task. For instance, all the work on trajectories has a special impact on real scenarios, such as analyzing patterns of moving cars in real roads. In this line, more results are expected, and also some prototypes will be released to disseminate these results.
Another line of research is the possibility of exploiting similarities and repetitiveness of genome collections for biological problems, such genome alignments. %This will be the focus of the PhD thesis of Diego Díaz (UCHILE), who will be seconded to UDC and UH in the second reporting period. 

\noindent{\bf Data structures and algorithms for network analysis}
There were also important contributions in this work package, in general more focused on information retrieval. Secondments and events generated joint contributions such as:
\begin{itemize}
\item A new compact trip representation over networks~\cite{SPIRE16-TRIP}, including algorithms for navigational rule derivation and determining the effect of traffic signs on those road networks~\cite{PACIS16-NAV}
%\item A new compact trip representation over networks (SPIRE 2016), including algorithms for navigational rule derivation and determining the effect of traffic signs on those road networks (PACIS 2016)
\item An efficient parallel construction of wavelet trees on multicore architectures~\cite{KAIS17} and of planar graphs~\cite{WADS17}.
%\item An efficient parallel construction of wavelet trees on multicore architectures (Knowledge and Information Systems 2016) and of planar graphs (WADS 2017).
\item An efficient representation for solving aggregated 2D range queries on clustered points~\cite{IS16-2D}, which was latter generalized, obtaining an efficient representation of multidimensional data over hierarchical domains~\cite{SPIRE16-CMHD}. This last contribution was part of a PhD thesis successfully defended in the University of A Coruña~\cite{PHD17}.% (July 2017).
%\item An efficient representation for solving aggregated 2D range queries on clustered points (Information Systems 2016), which was latter generalized, obtaining an efficient representation of multidimensional data over hierarchical domains (SPIRE 2016). This last contribution was part of a PhD thesis successfully defended in the University of A Coruña (July 2017).
\item An efficient computation of adjacency-matrix multiplication taking advantage of the compression obtained by graph compression methods~\cite{DCC18-MUL}.% (DCC 2018). 
\item A succinct data structure for self-indexing ternary relations~\cite{JDA17-TER}, which can be applied to RDF datasets representing information from biological fields, or exploited for any information retrieval task.
%\item A succinct data structure for self-indexing ternary relations (Journal of Discrete Algorithms, 2017), which can be applied to RDF datasets representing information from biological fields, or exploited for any information retrieval task.
\item A compressed representation of dynamic binary relations, in particular of graphs and networks that change over time~\cite{IS17-BIN}, which can be exploited to dynamic graphs and networks from any field.
%\item A compressed representation of dynamic binary relations, in particular of graphs and networks that change over time (Information Systems, 2017), which can be exploited to dynamic graphs and networks from any field.
\end{itemize}
During the second reporting period, we expect to obtain more contributions in the field of bioinformatics. For instance: designing a compact data structure to store the repeat graph, used for genome assembly, or using centrality measures over complex biological networks, such as those representing interactions among proteins, to obtain better representation of those networks. Moreover, there are big problems when visualizing complex networks. We expect to develop tools that help navigate and analyze biological trees or graphs using compact space, being time–efficient.

\section{Expected results and potential impact}
BIRDS will lead to results in the shape of products that have an impact on the research community and can be potentially transferred to the market as complete tools, libraries or frameworks for bioinformatics tasks or in other fields of IR. Some of the potentially exploitable results are:
\begin{itemize}
\item Compressed representations of repetitive sequences and repetitive collections. Compression and indexing of repetitive sequences provide the basis for basic search tools and also building blocks to run complex sequence analysis algorithms in reduced space. Preliminary results are already available as scientific publications. Repetitive sequences and collections arise not only in biological data, but also in many other areas of IR. These data structures and algorithms can potentially be exploited as tools in document-based databases or digital libraries and other document repository software, leading to overall improvements in space usage and query efficiency of software tools. 

\item Compressed representation of trajectories. Preliminary results have already been obtained regarding the compact representation of moving objects. The representation of trajectories is an open problem, especially in the field of geographical information retrieval. The storage of positions and trajectories of cars, ships, people, etc. is a usual requirement in many systems, which need to manage very large amounts of data. Efficient compression of trajectories is an active research field, but most of the solutions for efficient representation are limited to the research community. This will be exploited as a tool inside larger management tools, for companies that manage mobility workers, or container ships. In most cases, this kind of information is stored and transmitted uncompressed, leading to inefficiency. 

\item Compressed representation of multidimensional data. New algorithms and data structures for the self-indexed representation of multidimensional data provide the basis for advanced information retrieval over aggregated data. Several preliminary research results have already been obtained in this topic, which could be integrated and benchmarked to obtain a prototype compression tool. Also, the integration of compressed representations of multidimensional data and compressed representations of trajectories would provide a better solution for the use cases defined in the previous point.

\item Compressed representations of binary and ternary relations, graphs and networks. The representation of binary/ternary relations has many applications in the field of information retrieval, and can be applied in practice as a building block for network analysis and for the representation of sequences in general information retrieval. Specific compressed representations of relations, graphs and networks have been obtained as preliminary research results in the project, and improvements are expected. Representations of graphs, binary and ternary relations arise in many research fields, and efficient data structures and algorithms for the storage of graphs are relevant in many areas.

\item New algorithms for sequence alignment/assembly, and sequence analysis. Sequence analysis and specific tasks such as sequence alignment or genome assembly are essential for bioinformatics research. Any prototype or tool that improves space or time efficiency over state-of-the-art solutions would be immediately applied to further research and would also be of great impact in the industry. 

\end{itemize}

\section{Conclusions}
The project has achieved most of the goals planned for the first periodic report regarding research, training and networking activities. The consortium has achieved relevant scientific contributions, created new collaborations and enhanced existing ones, and proposed and implemented well-attended training and networking activities.
 
Regarding the first research task (Algorithms for Sequence Analysis), the most important results were in the field of Bioinformatics, where a new time-optimal algorithm for the subproblem contig assembly of the genome assembly problem was proposed. In addition, there were also some advances on the variation calling problem, which allows detecting mutations of an individual with respect to a reference genome.
	 	 	
For the second research task (Compression and indexing techniques for repetitive data), the knowledge of the partners on both fields were combined. As important results, we can enumerate: new universal indexes for highly repetitive document collections or efficient representation of trajectories and event sequences. In addition, there have been some important advances on new compression and indexing techniques for repetitive data not published yet. They will be an important contribution to the field, with applications on both bioinformatics and information retrieval.

Finally, results of the third research task (Data structures and algorithms for networks analysis) include a new compact trip representation over networks, an efficient parallel construction for wavelet trees planar graphs, an efficient representation of multidimensional data over hierarchical domains, a succinct data structure for self-indexing ternary relations, which can be applied to RDF datasets representing information from biological fields, and a compressed representation of dynamic binary relations, in particular of graphs and networks that change over time, which can be exploited to dynamic graphs and networks from any field.
%\end{document}  % This is where a 'short' article might terminate

\begin{acks}
 This project has received funding from the European Union's Horizon 2020 research and innovation programme under the Marie Sklodowska-Curie grant agreement No 690941.
 
Some works were also partially supported by MINECO [grant numbers TIN2016-78011-C4-1-R; TIN2016-77158-C4-3-R]; and Xunta de Galicia (co-founded with FEDER) [grant numbers ED431C 2017/58; ED431G/01].
 
%[G]Falta UDC

\end{acks}

\bibliographystyle{ACM-Reference-Format}
\bibliography{sample-bibliography}

\end{document}